\documentclass[aps,prl,twocolumn,superscriptaddress,notitlepage]{revtex4-1}
\usepackage{amsmath,amssymb,amsfonts}
\usepackage{graphicx}
\usepackage[dvipsnames]{xcolor}
\usepackage{comment}
\usepackage[overload]{empheq}
\usepackage{upgreek}
\usepackage{lmodern,pdftexcmds}
\usepackage[sort&compress]{natbib}
\usepackage{hyperref}
\usepackage{soul}
\usepackage{multirow}
\hypersetup{
    bookmarks=true,         
    bookmarksopen=true,
    unicode=false,          
    pdftoolbar=true,        
    pdfmenubar=true,        
    pdffitwindow=false,     
    pdfstartview={Fit},    
    pdftitle={},    
    pdfauthor={},     
    colorlinks=true,       
    linkcolor=black,          
    citecolor=black,        
    urlcolor=black,           
    pdfborder={0 0 0}
}

\renewcommand{\bm}[1]{\boldsymbol{\mathbf{#1}}}

\begin{document}

\title{Compact Expansion of a Repulsive Suspension}


\author{Matan Yah Ben Zion}
\affiliation{School of Physics and Astronomy and the Center for Physics and Chemistry of Living Systems, Tel Aviv University, Tel Aviv 6997801, Israel}
\affiliation{Department of Machine Learning and Natural Computing, Donders Institute for Brain, Cognition and Behavior, Thomas van Aquinostraat 4, Nijmegen, 6525GD, the Netherlands}
\author{Naomi Oppenheimer}
\email{naomiop@gmail.com}
\affiliation{School of Physics and Astronomy and the Center for Physics and Chemistry of Living Systems, Tel Aviv University, Tel Aviv 6997801, Israel}

\date{\today}

\begin{abstract}

Short-range repulsion governs the dynamics of matter from atoms to animals. Using theory,
simulations, and experiments, we find that an ensemble of repulsive particles spreads compactly
with a sharp boundary, in contrast to the diffusive spreading of Brownian particles. Starting from
the pair interactions, at high densities, the many-body dynamics follow non-linear diffusion with a
self-similar expansion, growing as $t^{1/4}$; At longer times, thermal motion dominates with the classic
$t^{1/2}$ expansion. A logarithmic growth controlled by nearest-neighbor interactions connects the two self-similar regimes.

\end{abstract}


\maketitle

Suspensions are everywhere --- from the ink we (used to) write with, the soy milk we drink, and the drugs we consume to the very structure of most living systems. 
Life builds on repulsive interactions in keeping microscopic particles suspended.
More often than not, the particles are charged, and electrostatic interactions are screened by the presence of ions in solution 
\cite{crocker1998interactions,likos2001effective, 
hu2021electrohydrodynamic}. Such is the case for charged proteins in a membrane \cite{sieber2007anatomy, oppenheimer2019rotating}, vesicles in suspension \cite{israelachvili2010direct, vlahovska2019electrohydrodynamics}, droplets in microlfuidic devices \cite{link2006electric}, water purification, plasma physics, and high charge-density batteries \cite{crow1975expansion,freidberg2008plasma, Li2018}. In other cases, particles are not strictly charged, yet are repelled by short-range forces, e.g., globular polymers, colloidal particles coated by a shell, or vortex cores in type II superconductors ~\cite{Grosberg1994,Milner1988,mladek2008multiple,Young2013, lacour2019influence, vieira2016general}.

Short-range repulsion is crucial for interactions spanning a wide spectrum of sizes and dynamics. These include the packing and flow of granular materials like sand~\cite{Jaeger1996}. It is also used to describe biological systems, elucidating the motion of bacterial colonies~\cite{baskaran2009statistical,Bechinger2016a}, the collective behavior of insect swarms~\cite{Ko2022, Peleg2018}, or the coordinated movement of vertebrates in herds, schools, flocks~\cite{Reynolds1987}, as well as for human crowds~\cite{Helbing2005, Bacik2023, Silverberg2013}. Furthermore, short-range repulsion is often utilized in modeling non-equilibrium many-body systems~\cite{redner110structure, solon2015pressure, yeo2015collective, tociu2022mean}, in information theory~\cite{Martiniani2019}, and in swarm robotics~\cite{BenZion2023}.

In what follows, we consider the expansion of a suspension of particles with repulsive, short-range interactions that dominate over thermal diffusion. We find that when the interaction has a typical decay length, the suspension expands compactly --- the concentration vanishes identically outside a core of finite size. Compact profiles are found in diverse physical systems, including gas diffusion through porous medium~\cite{barenblatt1952porous, pattle1959diffusion}, thin films~\cite{stone2002thinFilms,leal2007advanced}, and even in population dynamics~\cite{zhao2024traveling}. A family of compact solitons (called compactons) were found as solutions to a generalization of the Korteweg-De Vrie (KdV) equation ~\cite{rosenau1993compactons}. These systems were modeled using a continuum, hydrodynamic description characterized by phenomenological parameters. In this Letter, we show that the expansion of a dense suspension of particles with short-range repulsion follows two athermal regimes. The coarse-grained description of the microscopic pair-potential leads to a non-linear diffusion equation with a compact solution. We find under what conditions the continuum description breaks down, leading to a second regime.

\begin{figure}[tbh]
\centering
\includegraphics[scale=0.18]{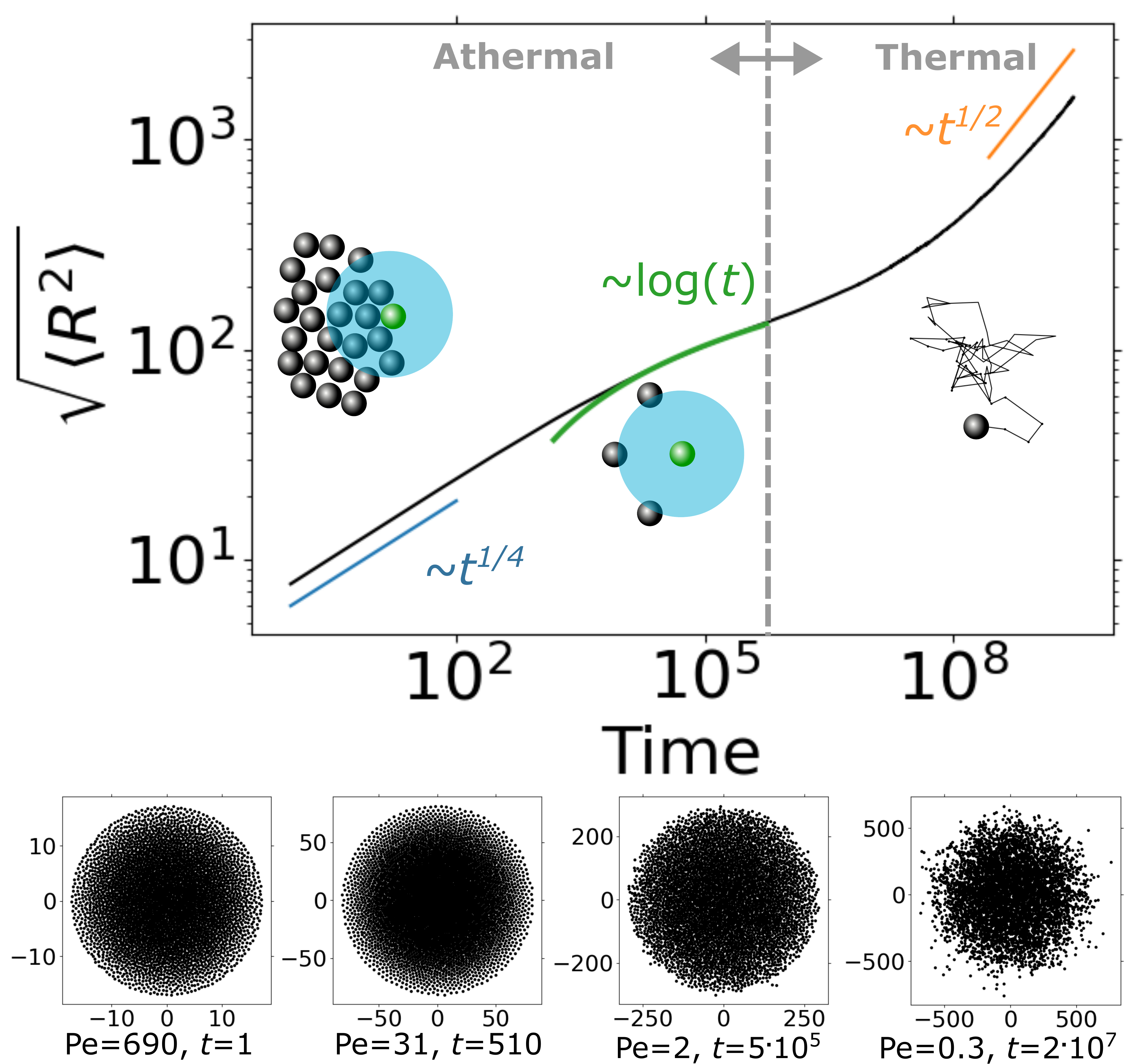}
\caption{
Simulation of 4000 particles starting from a dense random distribution.  At early times, the P\'eclet number is $\sim 4000$, and the dynamics are in the athermal regime. Each particle interacts with many others, and we see the $R\sim t^{1/4}$ scaling predicted by the self-similar solution. At intermediate times, each particle interacts only with its nearest neighbors, and there is a transition to the semi-sparse limit with a logarithmic dependence on time. At long times, when the P\'eclet number is smaller than one, the dynamics are governed by Brownian motion with the characteristic $t^{1/2}$ scaling, and the behavior is no longer compact, as can be seen from snapshots at the bottom.}
\label{figRadiusLongTimes}
\end{figure}

As we outline below, at high densities, the time evolution and distribution of the density field $n({\bf r,t})$,  are determined by a non-linear diffusion equation, stemming from particle-particle interactions, leading to a concentration-dependent diffusion of the form \cite{martzel2001mean, vieira2016general}
\begin{equation}
\frac{\partial n}{\partial t}  = \nabla \cdot \left[ (D_0 + \alpha n ) \nabla n\right], 
\label{eqNonlinearDiffusion}
\end{equation}
where $D_0$ is the thermal diffusion coefficient and $\alpha n$ acts as a density-dependent diffusion coefficient. Equation~\ref{eqNonlinearDiffusion} shows the relative significance of the thermal ($D_0$) and the athermal dynamics ($\alpha n$).
In our work we focus on the distinct nature of the athermal regime. For thermal diffusion, there is a Gaussian density profile while in the athermal case, the density profile is parabolic and is strictly zero beyond a maximal radius. Unlike classic diffusion where the radius of a drop grows as $\sqrt{t}$, a repulsion-dominated drop grows as $t^{1/4}$ (see supplementary video). The transition between the two regimes is dictated by nearest-neighbor interactions since particle separation is larger than the characteristic repulsive distance. For the ubiquitous exponential or screened electrostatic interactions, the asymptotic limits of the time evolution are given by 
\begin{equation}
R(t) \propto \begin{cases}
			t^{1/4} & \text{$n\gg 1$ \& ${\rm Pe} \gg 1$}\\
            \rm{log}(t) & \text{$n < 1$ \& ${\rm Pe} \gg 1$} \\
            t^{1/2} & \text{${\rm Pe} \ll 1$}
		 \end{cases},
\label{eqRadiusRegimes}
\end{equation}
where $n = \rho /\rho_c = l^2/L^2$ is the non-dimensionalized  density, $\rho \equiv 1/(\pi L^2)$, with $L$ being the typical distance between particles, and $\rho_c \equiv 1/(\pi l^2)$, with $l$ being the decay length of the short-range repulsion. 
The transition from the first to the second type of expansion occurs when $L = l$. 
The P\'eclet number (Pe), is the ratio between deterministic forces and thermal forces \cite{leal2007advanced}. Particles that are $10\;\mu\rm{m}$ or more (commonly found in food, cosmetics, and printing technologies) have negligible thermal diffusion and are expected to follow the athermal dynamics (high Pe). More generally,  the magnitude of the deterministic force depends on the mean inter-particle distance, $L$. Dense suspensions could be dominated by athermal dynamics even if individual particles have thermal diffusion. For example, from Eq.~\ref{eqNonlinearDiffusion} in the dense limit, $n\gg 1$, we can see that $D_0$ is negligible when $\alpha n \gg D_0$. Similarly, this is true when the mean distance between particles satisfies $L \ll \sqrt{F_0 l^3/k_bT}$, where $k_B$ is Boltzmann constant, $F_0$ is the magnitude of the force, and $T$ is temperature. This criterion for screened electrostatics can be alternatively written as $n\gg k_BT/2U_0$ (where $F_0l = 2U_0$). Note that in the over-damped dynamics discussed here, this limit is independent of the size of the particle. Thus, even small particles will show an athermal expansion at sufficiently short times when sufficiently dense (see potential applications for nanoparticles in SI).

Equation~\ref{eqRadiusRegimes} is the main result of this work, which is structured as follows: first, we analytically derive the two athermal regimes. Next, we compare analytic results with simulations and experiments. We show that at high densities ($n \gg 1$), where interactions go beyond nearest neighbors, discrete simulations are quantitatively consistent with the approximate analytical solution (as well as numerical integration of the mass conservation equation in the SM \cite{SM}); then we proceed to show that in the semi-sparse limit, where the dynamics are still athermal ($\rm{Pe} > 1$) but dominated only by nearest neighbor interactions  ($n\leq 1$) indeed the expansion follows Eq.~\ref{eqRadiusRegimes} as observed in both experiments of a charged colloidal suspension and discrete simulations.

\textbf{Governing Equations.}
We examine particles in the overdamped limit, where inertia is negligible, and the force, $\bf{F}$ and velocity $\bf{v}$,  are proportional through constant mobility, ${\bf v} = \mu {\bf F}$. The interaction can be due to any short-ranged, isotropic, repulsive force --- from sub-atomic Yukawa potential, through Pauli repulsion at the inter-atomic scale, screened Coulomb potential in an ionic solution, plasma, or even soft-core entropic repulsion in colloidal suspensions~\cite{Yukawa1935, Atkins2018, freidberg2008plasma,Grosberg1994,Chaikin1995}. Most of our results in the dense regime are generic, and for simplicity, we consider two-dimensional exponential interactions in the main text, $F(r) =  v_0 e^{-r/l}/\mu$. The SI has results for other forces. Our analysis does not apply to strictly hard-core repulsion.

\begin{figure}[tbh]
\centering
\includegraphics[scale=0.24]{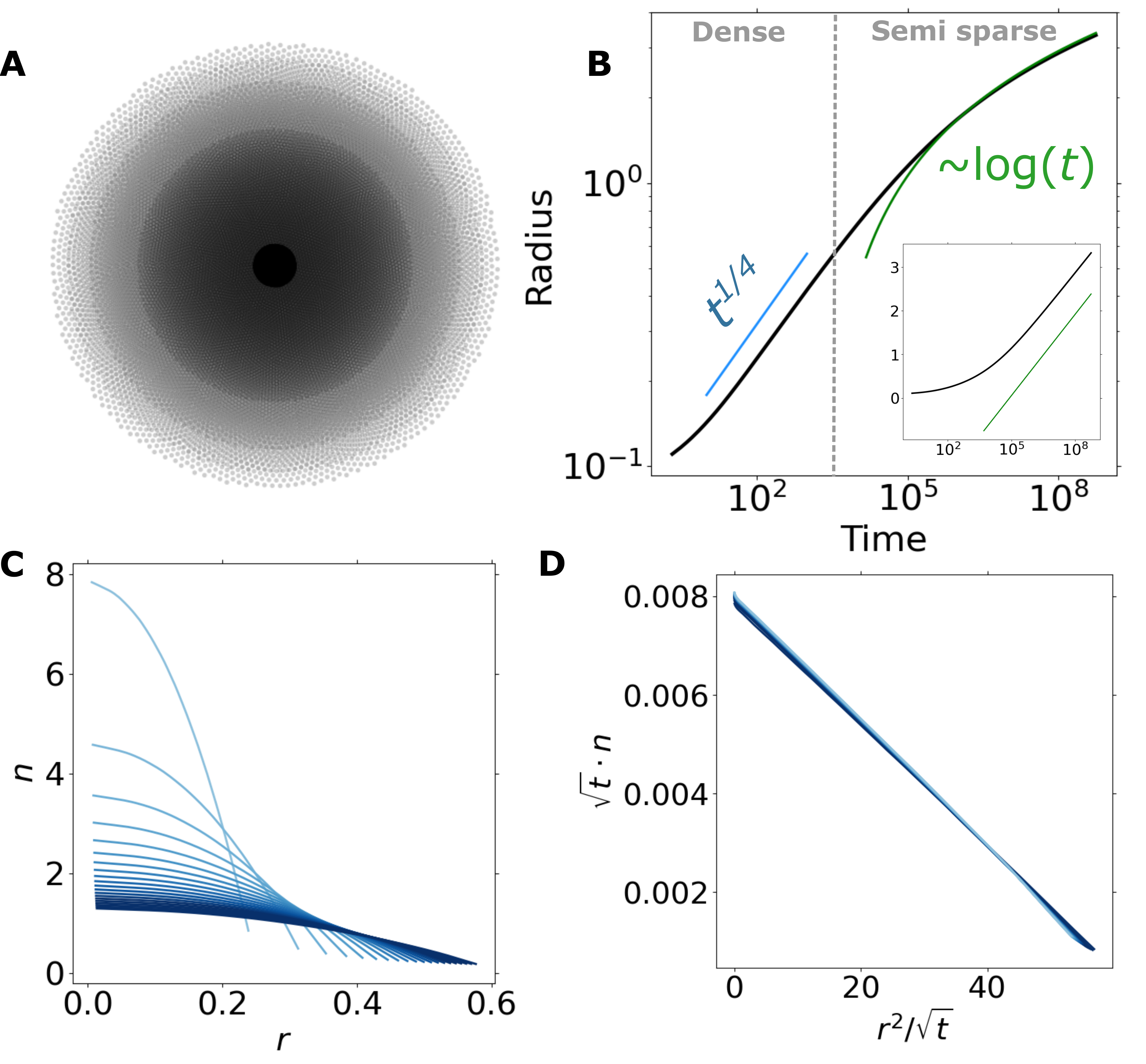}
\caption{
Results from a simulation of 10,000 athermal particles with exponentially repulsive interactions. We start from a high density and track the particles as they spread. (a) Snapshots of the simulations at different times ($t = 0, 5\cdot 10^3, 2\cdot 10^4, 3.5 \cdot 10^4, 5 \cdot 10^4$). (b) Radius as a function of time showing the $t^{1/4}$ scaling (in blue) in the dense limit, and as $\log(t)$ (green) in the semi-sparse limit. Inset shows the same data in a semi-log plot to better visualize the logarithmic regime. (c) Density as a function of radius in the dense limit for different times. The color goes from bright to dark as time progresses. (d) Re-scaled density $\sqrt{t}\, n$ as a function of $r^{2}/\sqrt{t}$ showing all the curves collapse to a single line as predicted by Eq.~\ref{eqSelfSimilar}.}
\label{figDensityVsR}
\end{figure}

To build intuition, let us examine two particles and then many. Two Brownian particles diffuse apart at a rate of $\sqrt{t}$. By contrast, two athermal, strictly repulsive particles separate as $\sim \log(t)$ since $dr/dt = 2v_0 e^{-r/l}$, where $v_0 $ is the magnitude of the velocity given by $v_0 = \mu F_0$. In an ensemble of many repulsive particles, each particle moves by the sum of the interaction from all particles. That is, the velocity of particle $i$ is given by, 
\begin{equation}
\mathbf{v}_i(\mathbf{r}_i)  = \mu \sum_j  F\left(\frac{r_{ij}}{l}\right)\hat{\mathbf{r}}_{ij} =  \sum_j v_0 e^{-r_{ij}/l}\hat{\mathbf{r}}_{ij},
\label{eqRepulsiveV}
\end{equation}
where $\mathbf{r}_{ij} = \mathbf{r}_i-\mathbf{r}_j$, $r_{ij} = |\mathbf{r}_{ij}|$, and $\hat{\mathbf{r}}_{ij} =\mathbf{r}_{ij}/ r_{ij}$ . Discrete simulations of Eq.~\ref{eqRepulsiveV} with the addition of a small Brownian noise term $\boldsymbol{\xi}$ (with $\langle \boldsymbol{\xi} \rangle = \boldsymbol{0}$  and $\langle \xi^2 \rangle \neq 0$)
show that the radius of an ensemble of $N$ repulsive particles grows as $\langle \sqrt{R^2}\rangle  \sim t^{1/4}$ in the dense limit ($n\gg 1$) and as $\log(t)$ in the semi-sparse limit ($n\leq 1$) , see Fig.~\ref{figRadiusLongTimes}. 
Initially, interparticle interactions dominate, ${\rm Pe} \sim 4000$, and thermal diffusion is negligible. As the suspension spreads, and its density decreases, interparticle interactions become weak. When ${\rm Pe} < 1$, Brownian motion dominate and the growth is of $t^{1/2}$. In the rest of the work we take $\xi=0$ and focus on the athermal limit.

\textbf{Analytic Results in the Dense Limit.}
In the dense limit, $n\gg 1$, we follow \cite{felderhof1978diffusion, martzel2001mean, vieira2016general, bruna2017diffusion} in coarse-graining the velocity to derive a diffusion equation. 
The procedure is analogous to a Fokker-Planck expansion with a mean-field closure . We start with the mass conservation equation for the number of particles 
$\frac{\partial n}{\partial t} + \nabla \cdot (n {\bf v}) = 0$,
where 
$\rho({\bm r(t)}) = \sum_i \delta({\bm r(t)} - {\bm r_i(t)})$. 
We turn to find the coarse-grained velocity, ${\bf v} ({\bf r}(t))$.
Since the interactions are purely repulsive, the velocity field is of the form ${\bf v}({\bf r}) = v(r) \hat{r}$. In the limit of a continuous density of particles, Eq.~\ref{eqRepulsiveV} becomes 
\begin{equation}
\label{eqRadialV}
{\bf v}({\bf X}) = \mu \rho_c \int_V n({\bf Y}) \frac{{\bf X}- {\bf Y}}{|{\bf X} - {\bf Y}|} F\left(\frac{|{\bf X} - {\bf Y}|}{l}\right)d^d{\bf Y},
\end{equation}
where $d$ is the dimension, and $V$ the volume of the drop. 
Combined with the mass conservation, the two equations can be solved numerically without approximations, as done in the SI. 

For particles away from the edge of the suspension, $|R-X|\gg l$, we can extend the integration boundaries to the entire space and perform a multipole expansion of Eq.~\ref{eqRadialV} in the density, giving the Taylor series, $
n( {\bf X}+ l {\bf s}) \approx n({\bf X}) + l {\bf s} \cdot \nabla n({\bf X}) + \dots$, where we have changed variables to a normalized distance ${\bf s}~=~({\bf Y} -{\bf X})/l$. In 2D polar coordinates ${\bf s} = s \hat{s} = s(\cos\theta, \sin\theta)$, ${\bf X} = r' (\cos\phi, \sin\phi)$, with $\theta, \phi \in [0,2\pi)$ and $s, r' > 0$.
By symmetry, the first term of the moment expansion vanishes after integration in Eq.~\ref{eqRadialV}. The remaining leading term in the velocity is the concentration gradient,
\begin{equation}
{\bf v}({\bf X}) 
\approx - \frac{l \mu}{\pi} \nabla n({\bf X})\cdot \int   \hat{s} \hat{s} d\Omega \int s^d F(s) ds
= -\alpha \frac{\partial n}{\partial r} {\hat r},
\label{eqVapprox}
\end{equation}
where $d\Omega$ signifies angular integration.
In both screened Coulomb and exponential repulsion in 2D $\alpha = 2 v_0 l$ (values for other cases in 2D and 3D are given in the SI). 

The full non-linear diffusion equation (Eq.~\ref{eqNonlinearDiffusion}) is found when plugging Eq.~\ref{eqVapprox} in the mass conservation equation,  giving an effective diffusion coefficient that linearly increases with density, $D = \alpha n(r)$. In 2D polar coordinates,
\begin{equation}
\label{eqDiffusionEquation}
\frac{\partial n}{\partial t} - \frac{\alpha}{r} \frac{\partial}{\partial r} \left(  r n \frac{\partial n}{\partial r} \right) = 0. 
\end{equation}
This equation is identical to the effective porous media equation \cite{barenblatt1952porous} but derived from the microscopic details of the pair interaction.
Self-similar solutions are given by dimensional analysis of Eq.~\ref{eqDiffusionEquation} (here we follow Refs.~\cite{leal2007advanced, stone2002thinFilms}): We start by assuming a solution of the form 
$n = A t^{\gamma} f(B r/t^{\beta}) = A t^{\gamma} f(\eta)$. We can further link $\gamma$ and $\beta$ by demanding that the total number of particles, $N$, is independent of time,
giving $\gamma = -2 \beta$. 
Placing $n$ in Eq.~\ref{eqDiffusionEquation}, we find 
$\beta = \frac{1}{4}$, and $B^2 = \frac{1}{8 A \alpha}$. The equation for the self-similarity function is 
$2f + \eta f' = -\frac{1}{2\eta} \frac{d}{d \eta} \left(\eta f  f' \right)$,
whose solution is parabolic, such that the concentration is
\begin{equation}
\label{eqSelfSimilar}
n = \frac{A}{\sqrt{{}t}}\left( 1 -\eta^2 \right) = \frac{A}{\sqrt{{}t}}\left( 1 - B^2 \frac{r^2}{\sqrt{t}} \right).
\end{equation}
Finally, the prefactor, $A$, is  determined from the total number of particles, 
giving $A = \sqrt{3 N/(8 \pi \rho_c \alpha)}$.
Note that the concentration of particles is \textit{strictly zero} beyond $ B r = t^{1/4}$, meaning that the drop is compact.  A similar calculation in 3D leads to $n = A t^{-3/5} f( Br/t^{1/5})$.

\textbf{Analytic Results in the Semi-Sparse Limit.}
When the average distance between particles is larger than the decay length of the repulsive force, $l$, we can assume only nearest neighbors contribute to the interaction, and the discrete nature of the suspension cannot be ignored. In such cases, we cannot use Eq.~\ref{eqVapprox} to find the density as a function of time. However, we can still approximate the radius of the drop as it spreads by considering the velocity of particles at the edge. Due to the repulsive interactions, the arrangement of particles is roughly hexagonal as verified in the simulations. We can assume a particle at the edge of the drop has three equally spaced nearest neighbors. Due to the isotropic nature of the interactions we can consider any particle. We take the particle positioned at $\mathbf{r} = R\hat{x}$ which moves with velocity
$\mathbf{v}\left(R\hat{x} \right) = \frac{dR(t)}{dt} \hat{x} = 2 v_0 e^{-R\sqrt{\pi/N}/l} \hat{x}$,
where $R\sqrt{\pi/N}$ is the average distance between particles in the ensemble. The approximate radius is simply found by integrating the velocity 
\begin{equation}
R(t) =  \sqrt{\frac{N}{\pi}} l \log(t/t_0+c), 
\label{eqRadiusEdge}
\end{equation}
with $t_0 = \sqrt{N/\pi} l /2 v_0$ and $c = \exp{(\sqrt{R_0^2\pi/l^2N})}$ where $R_0$ is the initial radius. We next demonstrate the validity of this result in simulations and experiments of a suspension of charge-stabilized colloids in deionized water. 

\begin{figure}[tbh]
\centering
\includegraphics[height=0.25\textheight]{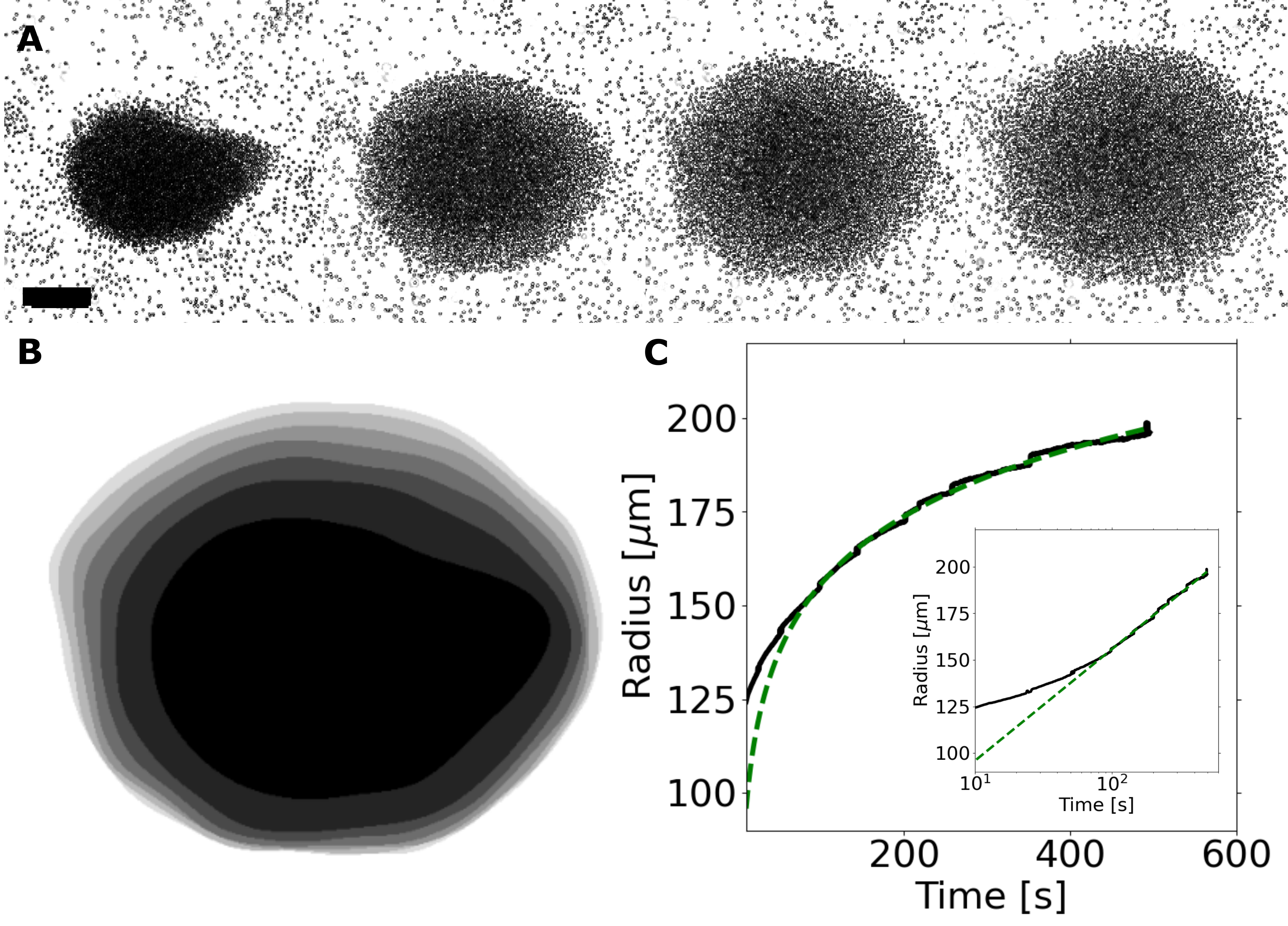}
\caption{Experimental results. 
(a) Snapshots of a dense suspension with $N\approx 3,000 $ particles show a compact expansion, with particles spreading due to screened electrostatic repulsion. Scalebar $100\;\mu\rm{m}$ (b) Overlapping figures with a color threshold (c) Drop radius as a function of time. The radius shows a logarithmic dependence (green dashed line) as predicted by the theory for a semi-sparse suspension.}
\label{figExperiment}
\end{figure}

\textbf{Discrete Simulation in the Dense Limit.}
We ran simulations of 10,000 particles with short-ranged exponential repulsion. We start from a random configuration in a circle of size $R$, ensuring that $n\gg 1$, and let the system evolve over time using a $5^{\rm th}$ order Runge-Kutta scheme. As the drop evolves, it spreads, such that $R = R(t)$. We find the local density of particles by using Voronoi tessellation and calculating the area of each cell, $A_{\rm cell}^i$~\cite{Oppenheimer2022}. The density is given by $\rho(r) = 1/A_{\rm cell}^i$ and $n(r) = \rho/\rho_c$. The upper left panel in Fig.~\ref{figDensityVsR} shows overlayed snapshots from the simulation at different times. Figure~\ref{figDensityVsR}B  shows the radius of the drop as a function of time. After a short transient, the radius follows the expected power-law of $R \propto t^{1/4}$. The density profile as a function of the radius of the drop, $r$, at different times, is presented in Fig.~\ref{figDensityVsR}C. And lastly, the bottom right panel shows the re-scaled density $\sqrt{t} \, n$ versus $r^2/\sqrt{t}$ in the dense regime. Note how all the curves collapse to a single straight line according to the scaling of Eq.~\ref{eqSelfSimilar}.

\textbf{Experiments of the Expansion of a Colloidal Suspension in the Semi-Sparse Limit}. 
We tested experimentally the expansion of a colloidal suspension. Our findings indicate that despite the presence of thermal motion of the individuals, the collective dynamics adhere to an athermal compact expansion. To achieve this, we used optical tweezers to concentrate the particles, following which we turned off the light and monitored the spreading of the colloidal drop (see Fig.~\ref{figExperiment}). Most commonly, optical tweezers have the laser light first enter the objective rear lens, coming to a tight focus at the imaging plane~\cite{Ashkin1970}. This creates a strong yet small trap that can typically host a single colloidal particle ($\sim1\;\mu\rm{m}$). To make a trap that can corral many particles, we built a custom optical setup by using a nearly collimated laser beam that first passes through the sample, and only then enters the objective through the collecting lens~\cite{BenZion2020, BenZion2022,Modin2023} (see SM for further details \cite{SM}).
We used $d = 3\;\mu \rm{m}$ carboxyl-functionalized colloidal particles suspended in deionized water with no added salts. The Debye screening length is expected to be between $\lambda_D \approx 0.2-0.6\;\mu\rm{m}$ \cite{okubo2000suspension,Behrens2000, dobnikar2004three,Behrens2000}. Using DLVO theory we estimate that when separated by a single radius, two charge-stabilized spherical particles with a screened-Coulomb interaction experience a repulsive potential of more than $100 \, k_BT$ (see SM \cite{SM}).

When the laser is turned on, particles are softly attracted into the region of a higher optical field, collecting approximately $N\approx 6000$ particles (see Fig.~\ref{figExperiment}A). Particles are packed at an effective area fraction of 0.9. Note that despite being in a  high filling fraction, the expansion is expected to follow the semi-sparse limit described in Eq.~\ref{eqRadiusRegimes} since $l\lesssim d$ and it is above the diffusive limit. Once the beam is turned off, the suspension starts to spread. During the initial minutes of spreading ($\sim 12$ min), the suspension remains compact with a sharp boundary (see supporting Movie). The velocity of the edge of the drop is initially $U \sim 4\;\mu$m/s, giving a P\'eclet number of ${\rm Pe} \sim 100$, such that the dynamics are indeed governed by inter-particle forces. The velocity decreases with density and towards the end of the experiment ${\rm Pe} \sim 1$, where the contribution of thermal diffusion becomes significant. We measure the size of the drop in the compact expansion regime by thresholding the movie (Fig.~\ref{figExperiment}) and extracting the radius of the drop at each frame. We find that as predicted by Eq.~\ref{eqRadiusRegimes}, the suspension expands logarithmically. By fitting the plot to a logarithm, we can approximate the number of particles ($\sqrt{N/\pi} l \approx 26 \;\mu m$). We find $N \approx 5900$, with $l = 0.6 \mu$m consistent with visual approximations. 

\textbf{Discrete Simulation in the Semi-Sparse Limit}.
Running simulations of 10,000 particles with exponential repulsion (or screened Coulomb repulsion in the SI) in the semi-sparse limit, results in a logarithmic growth of the drop's radius as a function of time.
Figure~\ref{figDensityVsR} shows a transition from $R\sim t^{1/4}$ at early times, as predicted by the self-similar solution, to $\sqrt{N/\pi} l \log(t)$ as predicted in the semi-sparse limit, Eq.~\ref{eqRadiusEdge}. Even though the analytic arguments made rough assumptions, namely, taking the average distance between particles as a measure of spacing, the coefficient of the logarithm is correctly predicted. In our case $l = 0.005$ giving $\sqrt{N/\pi} l = 0.28$.

In this work, we identified and characterized the athermal compact expansion of a repulsive suspension. We presented an analytical theory that captures the microscopic origin of the compact expansion and verified its different limits in both simulations and experiments. We identified two regimes where a collection of repulsive particles exhibits subdiffusive dynamics. (1) A dense regime where interactions go beyond nearest neighbors,  the radius expands as a power law in time ($R\propto t^{1/4}$), and the density profile is self-similar. (2) A semi-sparse regime where interactions are dominated by nearest-neighbor interactions. For exponential or screened electrostatics, the radius spreads logarithmically with time ($R\propto \log t$). The crossover between the two regimes occurs when the distance between particles becomes smaller than the decay length of the repulsive potential. Particles also interact hydrodynamically through the surrounding fluid, but the flow field is expected to be modest given that the suspension is near a solid boundary \cite{blake1974fundamental}, and is charged \cite{long2001note}.
Our analysis applies to cases where inter-particle interactions dominate over diffusion. These are applicable in both inherently athermal ensembles, such as granular matter and large organisms, but also for dense microscopic ensembles, as can be found throughout biology, and may serve to guide the design and synthesis of engineered colloidal matter.
\nocite{Allan2014}

\textbf{Acknowledgments.}
We thank Michael Shelley, Haim Diamant, Emir Haleva, and Philip Rosenau for insightful discussions. This work was supported by the Israel Science Foundation, grant number 1752/20.

\bibliographystyle{apsrev4-1}
\bibliography{compactRefs} 

\end{document}